# Life History Evolution in the Context of K Selection
## Solution of the Age-Structured Logistic Equation

Joshua Mitteldorf
Dept of Ecology and Evolutionary Biology
University of Arizona
Tucson, AZ 85721

josh@mathforum.org

## Abstract

$r$-selection refers to evolutionary competition in the rate of a population's free exponential increase. This is contrasted with $K$-selection, in which populations in steady-state compete in efficiency of resource conversion. Evolution in nature is thought to combine these two in various proportions. But in modeling the evolution of life histories, theorists have used $r$-selection exclusively; one reason is that there has not been a practical algorithm for computing the target function of $K$-selection. The Malthusian parameter, as computed from the Euler-Lotka equation, is a quantitative rendering of the $r$ in $r$-selection, computed from the fundamental life history variables mortality and fertility. Herein, a quantitative formulation of $K$ is derived in similar terms. The basis for our model is the logistic equation which, we argue, applies more generally than is commonly appreciated. Support is offered for the utility of this paradigm, and one example computation is exhibited, in which $K$-selection appears to support pleiotropic explanations for senescence only one fourth as well as $r$-selection.

# Life History Evolution in the Context of K Selection
## Solution of the Age-Structured Logistic Equation

**Introduction**

When competing populations invade an empty niche, fitness is well-measured by the population's proportional growth rate. When stable populations in steady-state compete in a quiescent environment, the controlling factor is not growth rate but efficiency of resource conversion, as measured by steady state population density. These two limiting cases are frequently contrasted as *r-selection* and *K-selection*. Both are thought to be important in nature, in varying proportions.

The *r* in *r-selection* comes from the theory of evolution of life histories. Fisher (1930) referred to this measure as the *Malthusian parameter*, and computed it from fertility and mortality curves using the Euler-Lotka equation. The *K* in *K-selection* has quite a different provenance; it comes from the logistic equation for population growth, and represents the carrying capacity of an environment for a given phenotype.

The logistic equation does not explicitly account for life history characteristics. Thus the way in which life history parameters contribute to *r*-selection has been well-defined for a long time; but *K*-selection has not previously been analyzed in these terms. Stearns (1992, p 207) recognizes this as a source of theoretical bias, and calls for analysis to bridge the gap: "...we do not yet know what changes in life history traits to expect under *K*-selection. One reason is that we do not have equations that relate changes in life history traits to changes in *K* that hold under density-dependence. Remedying this deficiency is a challenge for future research on life history theory." Our object in the present analysis is to respond to this challenge.

**Fisher's Malthusian parameter**

Relating the proportional growth rate of a population to the characteristic functions that define life history is an old problem, originally solved by Euler (1760), rediscovered in the context of modern population genetics by Lotka (1907), and first applied by Fisher (1930). Surprisingly, the problem has no algebraic solution, and *r* must be defined implicitly by an integral equation. The parameter *r* was dubbed by Fisher the Malthusian parameter, and the equation from which it is computed is referred to as the Euler-Lotka equation:

$$\int e^{-rx} l(x) b(x) dx = 1 \qquad (1$$

$l(x)$ is a survivorship function, the proportion of a population remaining alive at age $x$, and $b(x)$ is a fertility function, the mean number of offspring produced per unit time at age $x$. If we imagine that we already know *r*, then we have a means of comparing the relative contributions to growth of offspring produced at different times: their value declines exponentially with time at the prescribed rate *r*. The *r* that we seek, then, is that rate which reduces the total reproductive value of all offspring to unity. The procedure is closely analogous to the financial concept of return on investment: in that realm, dividends are discounted to present value by the factor $e^{-rx}$, and the r.o.i. is defined as that rate $r$ which causes the sum of the discounted present values of all future payments to equate to the original investment.

(It is easiest to think in terms of asexual reproduction, but incorporation of sex is straightforward and has little impact on the population dynamics. For sexual species, it is usually sufficient to track population, fertility, and mortality of females only. Assuming that all females find fertilization, and if females constitute a proportion $p$ of all individuals, then $r$ is reduced by a factor of $p$. (Stearns, 1992, p 21; Charlesworth, 1994 p 4))

With computational techniques that have become commonplace, the numerical determination of $r$ is fast and straightforward. The algorithm is seeded with a first guess for $r$, which is used to evaluate the integral numerically. The difference between the computed result and unity is fed into a Newton-Raphson or equivalent algorithm for generating a next-closer value of $r$, and the procedure is iterated until the desired accuracy attains.

## The classical logistic equation for one and several species

To obtain an analogous prescription for computing $K$, we begin with the classical logistic model: $\dot{N} = aN - mN^2$ is the simplest equation for describing populations whose growth is limited by finite resources. $N$ represents population size and the dot denotes time derivative; $a$ is a birthrate and $mN$ is the mortality rate, assumed to rise linearly with crowding. For a single species, a small population expands exponentially with rate $a$, but as the effect of crowding is felt, the quadratic term grows faster than the linear term. The population tends asymptotically toward a steady state of size $N^* = a/m$, where the growth rate vanishes. Anticipating that $N^*$ will become the $K$ of $K$-selection, we write the logistic equation in the form

$$\dot{N} = aN - mN^2 = aN(1 - N/K) \tag{2}$$

The model is readily extended to analyze competition among several phenotypes, whose relationship to environmental resources is similarly limiting: for subpopulations $N_i$, both growth and mortality are proportional to $N_i$, but the crowding term is proportional to the total population $N_T = \Sigma N_j$. Hence $dN_i/dt = a_i N_i - m_i N_i N_T$. Steady state population levels $K_i$ may be defined for each population in isolation, and, as above, the equations may be written in terms of these variables rather than the $m_i$.

$$\dot{N}_i = a_i N_i - m_i N_i N_T = a_i N_i (1 - N_T/K_i) \tag{3}$$

It is a well-known characteristic of the multi-species logistic model that that species with the greatest steady-state population density $K_i = a_i/m_i$ evolves to fixation. This is because so long as the total population $N_T$ is less than the steady-state level for any one species, that species has a positive growth rate; but once the total population surpasses steady-state values $K_i$ associated with each of the other species, those other species suffer a negative growth rate. Thus fitness in this model does not depend on the reproductive rate *a* separate from *K*; this is the sense in which *K*-selection represents an opposite extreme to *r*-selection.

(The multi-species logistic model is sometimes written in a more general way, with multiple independent coefficients for the sensitivity of each species to the presence of the others (Pielou, 1977). Distinguishing the effect of each species' crowding on the fortunes of every other is an important theoretical refinement, but it complicates the task of defining *K*: In a universe sufficiently complex to allow for the possibility that A may defeat B in evolutionary competition while B defeats C and C defeats A, there can be no single-valued function representing fitness. In the present discussion, we will find sufficient interest and generality in the form (3), where the different species contribute identically to a single crowding variable, to which the separate species may be more or less sensitive.)

## The age-structured logistic equation

Fertility and mortality curves, $b(x)$ and $m(x)$, define a life history. The Euler-Lotka equation is most conveniently written in terms of $b(x)$ and $l(x)$, where survivorship $l(x)$ is related to mortality $m(x)$ by the identity $d \ln(l)/dx = -m(x)$, or, equivalently, $l(x) = \exp(-\int_o^x m(x')dx')$. In the spirit of the logistic equation, we take mortality to be proportional to population density. To make this dependence on population explicit, we write $m(x)$ as:

$$m(x) \to Nq(x) \tag{4}$$

The survivorship curve $l(x)$ can then be computed from the mortality curve just as before:

$$l(x) = \exp(-\int_o^x Nq(x')dx') \tag{5}$$

The steady-state value of *N* is our fitness measure, *K*. We seek an equation which expresses the steady state condition, and which can be used to compute *N*=*K* implicitly. Such an equation can be derived from the condition $R_0=1$. $R_0$, the total lifetime reproductive output, can be constructed as an integral of fertility $b(x)$ weighted by survivorship $l(x)$, as follows:

$$\int_o^\infty b(x)[\exp(-\int_o^x Kq(x')dx')]dx = 1 \tag{6}$$

*K* in this equation is the steady-state value of *N*, written this way to emphasize that it is our candidate fitness measure. Fertility $b(x)$ and mortality $Kq(x)$ are regarded as measurable phenomena, and Eqn. 6 can be used to determine *K* implicitly, just as the Euler-Lotka equation is used to determine *r*.

Charlesworth (1994) is well aware of density dependent effects. He discusses the merits of carrying capacity as a fitness measure. In several different contexts (pp 52, 124, 148, 184), he declares that density dependence makes $R_0$ a function of total population *N*, and that in principle the equation $R_0(N)=1$ can be solved for $N^*$, yielding a measure of fitness. (In Charlesworth's notation, the equation appears as $w(n)=1$.) However, he stops short of displaying a practical computational scheme by which this may be accomplished.

## A more general interpretation of the logistic equation

The assumptions that fertility $a$ is a constant and that mortality $m$ is simply proportional to crowding, with no constant term, appear at first to be severely limiting, and to relegate the logistic model to a didactic device, or stepping stone to a more realistic representation. However, the same differential equation may be derived from a far more general model: Consider a monomorphic population in which fertility $A$ and mortality $M$ are free functions of total population $N$. If the population attains a steady state, that must occur for a level $N=N^*$ at which the fertility and mortality functions cross; furthermore, if the steady state is a stable one, then the difference $(A-M)$ must have a negative first derivative at $N^*$. The logarithmic population growth rate $(A-M)$ may be represented as a Taylor series, expanded about the equilibrium level $N^*$:

$$\dot{N}/N = 0 + [A'(N^*) - M'(N^*)](N - N^*) + ... \qquad (7$$

If we identify the logistic constant $a$ with $[A'(N^*)-M'(N^*)]N^*$, and $m$ with $[A'(N^*)-M'(N^*)]$, then the form $\dot{N} = aN - mN^2$ is recovered exactly, under broad assumptions, in a region sufficiently close to steady state $N=N^*$.

We have, then, a re-interpretation of the classical logistic equation. Our equation takes the same form, but $a$ is not a pure fertility term, nor is $m$ purely mortality; rather, they are defined in terms of the steady-state population level and the population's rate of change as it approaches steady state.

Charlesworth (1994 p 52) displays this same Taylor expansion, but takes no note of the logistic form of the first order result.

## When does the logistic model correctly predict competitive outcomes?

The logistic model is a differential equation for population dynamics, but the result that concerns us here is the endpoint of those dynamics. Since we propose to use the logistic $K$ defined as $a/m$ as a measure of fitness, it is interesting to ask under what circumstances the comparison of $K_1$ with $K_2$ correctly predicts the outcome of two-species competition.

Fig. 1 illustrates a situation in which species 1 tends toward a higher steady-state population level than species 2. Growth rate is not necessarily a linear function of total population, but the logistic equations are derived from the Taylor expansion about the two zero-crossings, and these crossings are represented exactly as $K_1$ and $K_2$.

For the two populations in the illustration, species 1 may reliably be predicted to replace species 2. This is because, for total population levels between $K_1$ and $K_2$, species 1 will expand steadily in population while species 2 diminishes.

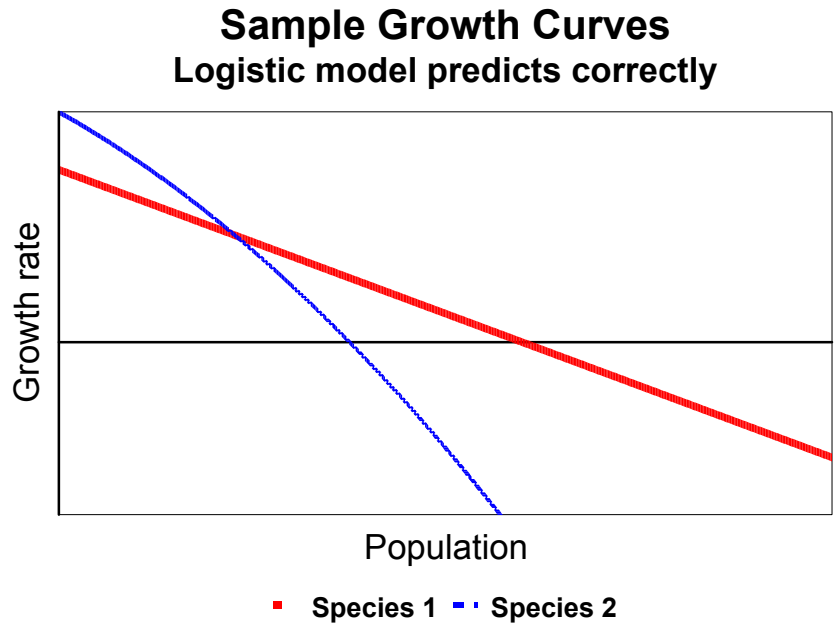

**Sample Growth Curves**
**Logistic model predicts correctly**

The logistic $K$ only fails as a predictor for growth curves that are so non-linear that they cross the axis more than once. The simplest such example is illustrated in Fig. 2. Species 2 has two metastable steady state population levels, and one unstable steady-state (where growth rate is zero, but the growth curve slopes upward). In this case, $K$ is not well-defined, and cannot be used as a predictive fitness measure.

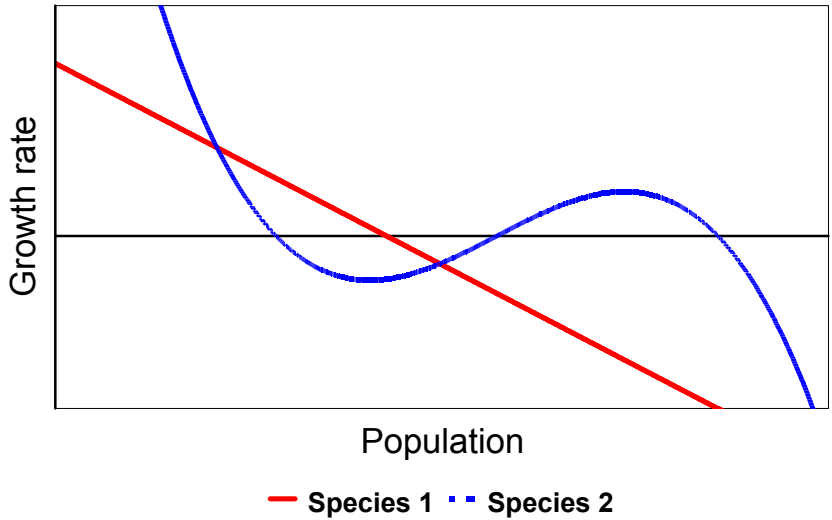

**Sample Growth Curves**
**Logistic model not reliable**

But so long as each species is characterized by a unique steady state population level $K_i$, the logistic model may be relied upon to correctly predict a competitive outcomes among them. The linearity of the logistic equation may distort the detailed prediction of population dynamics, but the accuracy of $K$ as a predictor of selective success survives intact.

### Effect of the generalized model on the integral equation for K

The generalized derivation does not change the form of the logistic equation (4), but only the interpretation of its symbols; the same is not true of the integral equation (6) from which $K$ may be computed. The new model projects a different form on the same condition ($R_0$=1) that we exploited above as defining steady-state conditions, and the steady-state population $K$. Following the logic of the Taylor expansion in Eqn. 7 above, we may allow for the possibility that fertility and mortality each have both a constant part and a part that varies linearly with population. Fertility and mortality are still permitted an arbitrary dependence on age $x$, and now their dependence on population $K$ is freed to be any first order polynomial. $b(x)$ in Eqn. 6 is replaced by $b_1(x)-Kb_2(x)$, and similarly $Nq(x)->q_1(x)+Nq_2(x)$. The result is

$$\int_o^\infty (b_1(x) - Kb_2(x))[\exp(-\int_o^x (q_1(x') + Kq_2(x'))dx']dx = 1 \qquad (8$$

Once again, Eqn. 8 is to be treated as an implicit equation for $K$, which is taken as a measure of fitness. All four functions $b_1(x)$ and $b_2(x)$, $q_1(x)$ and $q_2(x)$ are regarded as known measurables. In practice, measurement of $b_1$ and $b_2$ is equivalent to measuring fertility curves at two population levels, and $q_1$ and $q_2$ are similarly derivable from a pair of mortality curves.

## Special case: $q_2$=const implies equivalence of r and K

In the special case where $q_2$ is a constant, so that crowding mortality is independent of age, it turns out that *r* and *K* are equivalent measures. The conclusion depends on the assumption that sources of mortality are additive, so that total mortality can be expressed as above, $q_1(x) + Nq_2(x)$. This is true for the logistic case that we have considered, and more generally when crowding mortality is any monotonically increasing function of total population. Under these circumstances, any two life histories characterized by the same *r* also have the same steady state population level *K*.

Proof: *r* is by definition the growth rate of the population when it is small compared to K. *r* is computed from $q_1(x)$ alone, independent of $q_2$. If some age-uniform source of mortality *M* is imposed on the situation, then the shape of the age profile will be unaffected, but population growth rate will be reduced from *r* to *r* – *M*.

It follows that the mortality M needed to reduce the population's growth rate from *r* to zero is just *M=r*. In other words, if mortality *M* is a monotonic function only of total population, then population will rise until *M* just attains the value r. At that point, the population will no longer grow, so by definition this is the steady state population *K*. *K*, then, is that unique population level such that *M=r*, that is, the value of population for which crowding mortality = *r*. This is the sense in which *r* determines *K*.

## Numerical Validation

Adding age structure to the logistic model may also affect the reliability of *K* as a predictor of competitive success. Beginning with two competing populations, each having arbitrary age structures, two things happen simultaneously: One is that within each population, the age structure relaxes toward the steady-state distribution characteristic of its life history. The second is that the two populations compete, and one tends to replace the other. If the first process is permitted to proceed substantially to conclusion, then we know what the result of the second process will be: that species with higher steady-state population density will prevail, by the same argument that applies to the classical (non-age-structured) two-species logistic model. With small populations, and beginning with age distributions that are far from equilibrium, there is the possibility that the other species (with lower steady-state population density) may prevail, based on the short-term advantage of an age distribution that happens to be more heavily weighted toward youth and fertility.

In the computations reported below, the two different versions of the 2-species logistic equation were integrated numerically for sample life history functions $b_1(x)$ and $b_2(x)$, $q_1(x)$ and $q_2(x)$. For each case, life history parameters of one variety were adjusted and the computation was repeated until a stalemate between the two varieties was located; at this point the two values $K_a$ and $K_b$ were checked for equality, to verify whether our candidate fitness measure actually predicts the outcome of the competition. Of course, the values of $K_a$ and $K_b$ had also to be computed; the integrals over life history were evaluated repeatedly within a Newton-Raphson loop, which homed in on the implicit solutions. Computations were programmed in Borland Pascal, on a Windows-based desktop computer.

We were able to verify that equality of the two implicitly-derived values of $K$ corresponded to competitive stalemate when the two species were placed in competition. This we confirmed for a diverse set of life history functions $q_1$, $q_2$, and $b_1$, $b_2$. Fertility functions $b$ included biologically "reasonable" as well as unreasonable functions, with smooth dropoffs, finite discontinuities, short and long times to maturation. Mortality curves $q$ again included functions both smooth and discontinuous, rising, falling, and turning over. Numerical integration over age was performed with as few as 10 age bins and as many as 1000. The only disparities between the stalemate values of $K$ could be adequately explained by the finite difference approximation.

We argue above that this result was expected so long as populations are able to relax to steady-state age distributions before an irremediable advantage is attained by one side or the other. In the real world, this may be an important limitation, since one population may fall to zero before steady-state age distributions can be established. But our simulations modeled populations with floating-point variables, which never fall literally to zero; hence it was sufficient that distortion of age distribution curves from the competitive conditions not be so severe as to affect the competitive outcome. In practice, we were unable to construct a counterexample in which $K$ did *not* accurately predict competitive outcomes.

Note that this is a deterministic numerical integration, (not a stochastic model) equivalent to the infinite population limit in which we expect a fitness advantage to be absolutely determinative. Small populations are more realistic for some cases, but results depend so critically on a variety of initial specifications that generalization about this issue is reserved for another study. Our preliminary investigation indicates, however, that $K$ is likely to statistically predict the outcome of any 2-species competition whenever total population sizes are $\gtrsim 30$. For smaller populations, initial conditions and stochastic variation make $K$ less reliable as a predictor.

## Application

Understanding the evolution of senescence has been the most important and challenging application in the field of life history evolution. In this context, Fisher's $r$ has frequently been invoked, and less frequently actually applied in computations apropos to senescence. The following is offered as an example of the usefulness of our prescription for steady state population level as a measure of fitness.

Since it was formulated by Medawar (1952) and articulated by Williams (1957), antagonistic pleiotropy has been a leading candidate for a theory of evolution of senescence. The essence of the theory is that adaptations favoring early fertility will be selected, even if they entail systemic stresses that result in accelerated mortality late in life. Fisher's Malthusian fitness measure has

-9-

frequently been cited in a general way as justification for this paradigm: the structure of the formula makes it clear that fertility early in life makes a large contribution to $r$, and that mortality at later stages is discounted in importance for two reasons: first, because $e^{-rx}$ is already small at late times, and second, because the cumulative effects of earlier mortality has already reduced the affected population.

With $K$-selection, the qualitative argument for antagonistic pleiotropy remains valid, but only the second of the two reasons remains potent. As we see in the example below, the strength of the effect can be substantially diminished.

In our first application of the theory of $K$-selection on life histories, we compare the net contribution to fitness from enhanced fertility accompanied by senescence. A control population has constant mortality (no aging) and baseline fertility (also constant with age); a variant population enjoys a fertility level 10% higher, but suffers aging according to a Gompertz (1825) curve, such that mortality (after sexual maturity) increases as $e^{cx}$, where age $x$ is measured in units of the (constant) mortality rate ($d\ln(N)/dx$); $c$ is then adjusted until the varieties are equally fit under $r$-selection; the computation is then repeated under $K$-selection. For the former case, the two populations are equal competitors when the 10% fertility advantage is offset against mortality increasing at an exponential rate $c=3.56$. But, repeating the computation using $K$ for fitness, the same 10% enhancement in fertility can only offset a Gompertz rate of $c=0.94$. The implication is that, in this example, antagonistic pleiotropy is roughly one fourth as effective as a force for evolving senescence in $K$-selection as compared to $r$-selection.

The parameter details for this calculation are as follows: fertility was held constant at 11 and 12.1 for the two cases, measured in units where mortality=1. A reproductive maturity time of 0.1 was used for both parts of the computation, measured in units of the reciprocal mortality rate. For the second part of each calculation, the Gompertz mortality was taken to be unity for time less than maturity, and $e^{c(x-\alpha)}$ at later times, where $c$ is the Gompertz factor referenced above, and $\alpha$ is the age of sexual maturity. For the logistic computation, the steady-state population was computed to be $K=6.02$ in the constant mortality case, and $c$ was adjusted until the same result was achieved for Gompertz mortality after fertility had been enhanced by 10%.

Although this computation was based on a specific set of functions, we expect very generally that $K$-selection is less friendly to theories of senescence based on antagonistic pleiotropy than is $r$-selection.

## Summary

Fisher's Malthusian parameter was originally proposed as a fitness measure derived from life history functions, at a time when $K$-selection was a concept not yet in currency. Seventy years later, analysis of life histories is still based exclusively on $r$ and never $K$.

Recently, Benton and Grant (2000) have devised a simulation for comparing the predictive power of the standard fitness measures; their conclusion is that carrying capacity $K$ is by far the most robust of these, and they discuss the reasons that *r* is likely to be less relevant in realistic ecologies. This conclusion also is consonant with our biological common sense: the lavish diversity of adaptations which we observe in nature are directed at directed competition, at protection, at efficient use of available resources, and at every aspect of survival; only a fraction of this diversity is aimed at enhancement of the raw quantity of seed product.

The Euler-Lotka equation is not easily adapted to applications where one would most like to test it: numerical simulations and evolutionary theory of age-structured populations, where intense competition holds populations levels close to their steady-state values. We offer, in analogy to Euler-Lotka, a fitness measure that can be computed from life history variables and which is appropriate to situations in which *K*-selection predominates. We exhibit two implicit equations for computation of *K*, which is simply the steady-state population level. We have verified that solutions to these equations do indeed predict the outcome of evolutionary simulations where generalizations of the logistic equation are integrated numerically. We have noted two limitations to the applicability of our analysis: first, that it does not allow for separate coefficients to quantify the effect of each variety's population density on the mortality of the other; second, we have tested the predictive value of *K* for continuous population variables only, and have not explored the interesting but complex behavior of populations sufficiently small to be subject to random extinctions.

We have argued for the generality of the logistic paradigm: Assuming arbitrary dependence of mortality on crowding, the logistic equation may be constructed from the zero and first order Taylor coefficients in deviation from steady-state; the form of the classic logistic equation is recovered, but the two logistic constants must be interpreted as each containing information about both birth and death rates. We conclude that for populations that are sufficiently close to their steady-state densities, the logistic model and the general thrust of the present analysis are applicable. More generally, we argue that the model correctly predicts the result of evolutionary competition so long as several species share a common set of environmental resources such that each is characterized by a unique steady-state population level. For populations that are periodically curtailed from steady state levels, Fisher's *r* is one factor contributing to fitness, but resistance of each subpopulation to the source of the devastation is an equally important factor, so that detailed simulation is probably necessary to determine the outcome of any specific contest.

We offer one sample application, in which it appears that *K*-selection is only one fourth as effective as *r*-selection for explaining the evolutionary provenance of senescence via a pleiotropic link to fertility.

## Acknowledgments

Warren Ewens posed the problem and supplied the original inspiration for this work. Peter Taylor read an early draft and contributed to my understanding of the issues. Stephen Stearns gave me a context for the introduction and conclusions.